\documentclass[11pt,preprint]{aastex}

\usepackage{verbatim}
\usepackage{natbib}
\usepackage{amsmath}
\usepackage{amsbsy}
\usepackage{lscape}
\usepackage{epstopdf}

\slugcomment{Submitted to ApJ}
\shorttitle{Cloud infall into the Galactic supermassive black hole}
\shortauthors{Burkert, Schartmann et al.}

\begin{document}

\title{Physics of the Galactic Center Cloud G2, on its Way towards the Super-Massive Black Hole} 

\author{A. Burkert\altaffilmark{1,2,3}, M. Schartmann\altaffilmark{1,3},
C. Alig\altaffilmark{1,3}, S. Gillessen\altaffilmark{3}, R. Genzel\altaffilmark{3},
T.K. Fritz\altaffilmark{3} and F. Eisenhauer\altaffilmark{3}}

\altaffiltext{1}{University Observatory Munich, Scheinerstrasse 1,
D-81679 Munich, Germany} \altaffiltext{2}{Max-Planck-Fellow}
\altaffiltext{3}{Max-Planck-Institute for Extraterrestrial Physics, Giessenbachstrasse
1, 85758 Garching, Germany} 

\email{burkert@usm.lmu.de}

\newcommand\msun{\rm M_{\odot}}
\newcommand\lsun{\rm L_{\odot}}
\newcommand\msunyr{\rm M_{\odot}\,yr^{-1}}
\newcommand\be{\begin{equation}}
\newcommand\en{\end{equation}}
\newcommand\cm{\rm cm}
\newcommand\kms{\rm{\, km \, s^{-1}}}
\newcommand\K{\rm K}
\newcommand\etal{{\rm et al}.\ }
\newcommand\sd{\partial}
\newcommand\Topspace{\rule{0pt}{2.6ex}}
\newcommand\Botspace{\rule[-1.2ex]{0pt}{0pt}}

\begin{abstract}
The origin, structure and evolution of the small gas cloud, G2, is investigated,
that is on an orbit almost straight into the Galactic central supermassive black hole (SMBH). 
G2 is a sensitive probe of the hot accretion zone of Sgr A$^*$, requiring gas temperatures and 
densities that agree well with models of captured shock-heated stellar winds. Its
mass is equal to the critical mass below which cold clumps would be destroyed quickly by evaporation.
Its mass is also constrained by the fact that at apocenter its sound crossing timescale was equal 
to its orbital timescale. Our numerical simulations show that the observed structure and evolution of G2
can be well reproduced if it formed in pressure equilibrium with the surrounding in 1995 at a distance from
the SMBH of 7.6 $\times 10^{16}$ cm. If the cloud would have formed at apocenter
in the 'clockwise' stellar disk as expected from its orbit,
it would be torn into a very elongated spaghetti-like filament by 2011 which is not observed. This problem
can be solved if G2 is the head of a larger, shell-like structure that formed at apocenter. 
Our numerical simulations show that this scenario explains not only G2's observed kinematical
and geometrical properties but also the Br$\gamma$  observations of a low surface brightness gas tail that 
trails the cloud. In 2013, while passing the SMBH G2 will break up into a string of droplets 
that within the next 30 years mix with the surrounding hot gas and trigger cycles of AGN activity.
\end{abstract}


\section{INTRODUCTION}
The Galactic center is one of the most extreme and puzzling places of the Milky Way.
Harboring a supermassive black hole (SMBH) with a mass of M$_{BH} = 4.31 \times 10^6$ M$_{\odot}$
(Ghez et al. 2008; Gillessen et al. 2009, Genzel, Eisenhauer \& Gillessen 2010) at the position
of the radio source Sgr A$^*$ it is by far the closest and most ideal place to investigate
the physics of galactic nuclei and their activity cycles when gas accretes onto the SMBH. 
Most galaxies contain SMBHs that correlate well with various global
galactic properties (G\"ultekin et al. 2009; Burkert \& Tremaine 2010; Kormendy, Bender \& Cornell 2011). However,
only a small fraction is observed to be active at any time
(e.g. Heckmann et al. 2004, King \& Pringle 2007). The processes that lead to these
long phases of quiescence with no or little SMBH accretion, interrupted by
short periods of activity, are not well understood up to now.

The Milky Way's central SMBH currently is classified as inactive, despite observations of
irregular flickering events (Baganoff et al. 2001; Genzel et al 2003) that demonstrate that it is still accreting
small amounts of mass, sporadically. The 'X-ray echo' in the molecular clouds near Sgr A$^*$ might be a signature of such
an accretion event that occured a few hundred years ago with an energy
output of order $10^{39}$ erg/s (Sunyaev et al. 1993, Koyama et al. 1996, 2003, 2009, 
Revnivtsev, Molkov \& Sazonov 2006).
Evidence also exists of a major outburst 1-10 Myr ago. 
The two giant Fermi bubbles (Su, Slatyer \& Finkbeiner 2010), originating from the Galactic
nucleus and extending to 10 kpc perpendicular to the galactic disk are impressive signatures 
of such an active phase (Cheng et al. 2011). It is tempting to speculate that the origin of
the Fermi bubbles is related to the formation of the
$\sim$ 100 massive, young O- and Wolf-Rayet stars that have been found within the central 0.1 pc 
(Genzel et al. 2003; Ghez et al. 2005; Bartko et al. 2009). Many
of these stars orbit Sgr A$^*$ in two counter-rotating and inclined disks which probably 
formed from one or two massive gas clouds that fell into the Galactic nucleus $\leq$ 1-10 Myr ago. 
In addition, numerical simulations show
that a $10^5$ M$_\odot$ cloud, interacting with the SMBH would naturally form
a thin, dense accretion disk that cools and condenses into a disk 
of massive stars in its self-gravitating parts, outside the SMBH's Bondi
radius $R_B \approx 0.1$ pc  (Baganoff et al. 2003), in good agreement with the radius regime
of the observed stellar disks
(e.g. Nayakshin, Cuadra \& Springel 2007; Bonnell \& Rice 2008; Hobbs \& Nayakshin 2009; Alig et al. 2011). 
Alexander et al. (2011) however point out that a long-lived, gravitationally stable, optically thick
residual gas disk should be left behind within R$_B$, feeding the SMBH on timescales longer than
10 Myr (Nayakshin \& Cuadra 2005). Observations completely rule out such a 
compact disk which would be easily detectable in the mid-IR (Sch\"odel et al. 2011).
The puzzle of the missing gas disk is still poorly understood. It is however consistent with the 
quiescent, low-luminosity AGN state of Sgr A$^*$ (Loeb 2004) with very low Eddington accretion rates of
$\leq 10^{-6} \dot{M}_{Edd}$ (Bower et al. 2003).

Instead of cold gas, {\it Chandra} observations of the Galactic center have revealed
a tenuous, hot, ionized, X-ray emitting gas bubble (Baganoff et al. 2003) with temperatures of order 1-4 keV
that is believed to originate from the shock-heated strong stellar winds of surrounding massive stars 
(Melia 1992; Krabbe et al. 1991; Najarro et al. 1997). Quataert (2002, 2004) investigated
the structure of such a hot gas component. His models predict
that only a few percent of the total stellar mass loss, of order a few $\times 10^{-5}$
M${_\odot}$/yr, should by gravitationally captured within the Bondi radius and
accrete onto the SMBH, with the rest being driven out of the nucleus. This accretion rate is still
large compared to the values of $\leq 2 \times 10^{-7}$ M$_{\odot}$/yr,
as inferred from Faraday rotation measurements (Marrone et al. 2007).
Sazonov, Sunyaev \& Revnivtsev (2011) also point out that this accretion rate would lead to 
a X-ray luminosity of the SMBH that would be much higher than the observed
$\sim 10^{33}$ erg s$^{-1}$, which implies lower accretion rates of 
$\leq 10^{-7}$ M$_{\odot}$ yr$^{-1}$
(Bower et al. 2003). They note that the required reduction in hot gas density inside the Bondi radius
is in agreement with {\it Chandra} observations, if a substantial or even dominant fraction of
the X-ray emission is actually produced by
late-type main-sequence stars of the central stellar cusp.

The situation is even more complex. Detailed 3-dimensional numerical simulations of the stellar
wind interaction of the young massive stars in the Galactic center
(Cuadra et al. 2005, 2006; Nayakshin, Cuadra \& Springel 2007; Nayakshin \& Cuadra 2007; Martins et al. 2007;
Cuadra, Nayakshin \& Martins 2008) demonstrate that fast winds 
from Wolf-Rayet stars with velocities of $v_w \geq 1000$ km/s are shock heated to temperatures
$> 10^7$ K. In contrast, luminous blue variable stars with wind velocities of 300-600 km/s, when shocked, 
generate gas with temperatures of $10^6$ K that can cool fast and form cold, ionized $10^4$ K gas clumps and filaments
that fall onto the SMBH, leading to short bursts of activity. In this case, the accretion rate
of the SMBH, averaged over times of order a typical clump infall timescale, could be much larger
than estimated from its current inter-burst X-ray luminosity. The existence
of such a cold gas component would therefore have strong implications
for our understanding of the activity and growth the Milky Way's SMBH.

Strong observational evidence for a cold cloud component in the Galactic center 
has recently been reported by Gillessen et al. (2012). They detected a
dense, dusty and ionized gas clump which we will call G2 
(another cloud, G1, has been discovered earlier by Cl\^{e}net et al. (2005))
with a dust temperature of merely 550 K and a gas temperature of order $10^4$ K, 
representing a relatively cold droplet of ionized gas, embedded in the
diffuse $10^8$ K gas of the central hot bubble. 
The cloud's radius is resolved in the direction of its orbital motion but unresolved in the perpendicular direction. 
Gillessen et al. (2012) adopt an effective, spherical
radius in 2011.3 of 15 mas which corresponds to $1.875 \times 10^{15}$ cm.
For case B recombination and a mean molecular weight of $\mu = 0.6139$
the observed Br$\gamma$ luminosity implies a cloud density of $\rho_c \approx 6.1 \times 10^{-19} f_V^{-1/2}$ g cm$^{-3}$
with a corresponding cloud mass of M$_c \approx 1.7 \times 10^{28} f_V$ g or approximately 3 Earth masses. 
Here $f_V \leq 1$ ist the volume filling factor. In this paper we will investigate scenarios where the cloud is a 
compact region of cold gas and therefore adopt $f_V = 1$.
The observations (contours in Figure 5; see also Figure 2 in Gillessen et al. 2012)
indicate, that G2 might be the upstream compact head of a larger, diffuse distribution of cold gas.
The total mass of this system could be substantially larger than G2's estimated mass.

The cloud is approaching
Sgr A$^*$. Its orbit could be traced back for the past 10 years, allowing
a detailed orbital analysis which shows that G2 is moving on a highly
eccentric orbit. In 2013.5 the cloud will pass the SMBH at a pericenter distance 
of merely 3100 times the event horizon, corresponding to  $4.0 \pm 0.3 \times 10^{15}$ cm. 
The next two years will therefore
provide a unique opportunity to investigate directly the disruption of a cold gas clump
by its gravitational interaction with the SMBH and maybe even the onset of a new activity cycle, 
triggered by the accretion of gas onto the central black hole.
In fact, clear evidence for ongoing tidal velocity shearing and stretching has already been detected.

The existence of a tiny, cold gas cloud in the near vicinity of the SMBH, 
embedded in the hostile $\sim 10^8$ K gaseous environment, is surprising and raises 
numerous interesting questions. Where did this cloud come from
and where will it go? Why is it on such a highly eccentric orbit?
Which processes constrain the physical properties of G2, i.e. its size, mass,
density, temperature and geometrical shape. How many clouds like G2
are currently orbiting Sgr A$^*$ and how do they affect its activity and 
gas accretion rate?

In this paper we will investigate some of these questions,
focussing on a simplified prescription of the cloud's structure and evolution. 
More detailed numerical simulations will be presented in subsequent papers.
Section 2 discusses possible formation scenarios. Section 3 summarizes the orbital 
properties of G2 which have implications for its formation.
Section 4 investigates the various hydrodynamical processes
of interaction between G2 and its surrounding that constrain G2's origin, orbit
and evolution.
In Section 5 we show that G2 can in turn be used as a sensitive probe to explore
the thermodynamics of the diffuse gas component in the Galactic Center. Section 6
finally presents
a first set of hydrodynamical simulations that explore the evolution of a cold
gas cloud, moving through a hot, stratified medium within the gravitational potential
of the SMBH. The results are summarized  and conclusions are drawn in Section 7.

\section{FORMATION SCENARIOS}

Possible formation scenarios can be broadly
subdivided into two categories, the diffuse cloud scenario versus 
the compact source scenario. 

\subsection{The Diffuse Cloud Scenario}

The diffuse cloud model assumes that G2 is a cold gas clump,
embedded in and confined by the hot gaseous environment of the central bubble. For the densities and
temperatures of the diffuse hot gas in the Galactic Center, 
the cooling timescale of the hot gas is much longer than the dynamical timescale
(Cuadra et al. 2005; Quataert 2004). 
Cloud formation due to a cooling instability (Field, Goldsmith \& Habing 1969;
Burkert \& Lin 2000) of the hot gas can therefore be ruled out as a formation mechanism.
The most likely scenario is probably that G2 originates from the complex two-phase
medium in one of the two disks of young massive stars that surround the central bubble.
Shocked, fast winds of young massive stars with velocities of order
1000 km/s generate hot $10^8$ K  plasma, that partly is gravitationally capture
by the SMBH and forms the central hot bubble. In contrast, slow
winds from luminous blue variable candidates with velocities of 300-500 km/s when shocked
are heated only to $10^6$ K and subsequently cool fast, leading to the formation of cold,
bound cloudlets, embedded in the hot intercloud medium. Interestingly, the orbital plane
of G2 coincides with the clockwise disk of young, massive O and Wolf-Rayet stars 
and its apocenter distance agrees with the disks's inner edge (left panel of Figure 1).
This supports the notion that the cloud might indeed be shocked wind debris. The circular velocities
of the stars in the disk are of order 700 km/s while G2 started on a highly eccentric
orbit with a tangential velocity at apocenter of order 200 km/s (see Section 3). 
For typical wind velocities of 500 km/s and assuming a rotational velocity of
the star of order 700 km/s, clumps might 
then have rotational velocities that range between 1200 km/s upstream and 200 km/s downstream.
While the upstream clumps might be ejected from the Galactic center, downstream clumps would have just
the correct initial velocity to fall into the center on a highly eccentric orbit, like G2.
One of the implications of this model is that cold clumps with a large spectrum of masses
and orbital parameters should continuously be produced by this process
in the stellar disks with many of them falling into the
central region on a highly eccentric orbit. We will show however in the subsequent sections
that G2's mass is special in the sense that
clumps with masses larger than G2 might easily break up into smaller pieces while smaller clumps 
evaporate quickly.

\subsection{The Compact Source Scenario}

Within the framework of the compact source scenario, G2 represents
the visible diffuse gas atmosphere of an unresolved, dense object in its center
that might continuously loose gas (Gillessen et al. 2012; Murray-Clay \& Loeb 2012). 
The object might have formed $10^6$ yr ago in the stellar disk and was 
scattered into a highly eccentric orbit due to a close encounter with another star or
massive black hole. Gillessen et al. (2012) argue that
the gas cloud must be optically thin. In order to be invisible its central, compact object
has to be either hotter than $10^{4.6}$ K, emitting most of its light in the ultraviolet with
a low enough luminosity of $< 10^{3.7}$ L$_{\odot}$ or very cold. Gillessen et al. suggest
a compact planetary nebula as a possible candidate. Murray-Clay \& Loeb (2012) propose
a dense, protostellar disk, bound to a low-mass star. Another possibility might be an
evaporating low-mass star, brown dwarf or Jupiter-like planet. 

In order to distinguish between the different scenarios and determine
G2's nature it is important to find constraints that need to be met by any viable
model. In the following sections we will investigate characteristic properties
of G2's structure and orbit which are provide information about the properties
of the confining hot gaseous environment. We then focus on the diffuse cloud scenario in detail
using hydrodynamical simulations that follow the cloud on its way towards pericenter
and dispersal afterwards.

\section{ORBITAL PARAMETERS OF G2 AND ITS LOCATION OF BIRTH}

Measurements of the position and velocity of G2 tightly constrain its orbit (see Gillessen et al. 2012 for details).
Adopting a SMBH mass of $4.31 \times 10^6$ M$_{\odot}$ and a Sun - Galactic Center
distance of 8.33 kpc, the orbital eccentricity is e=0.9384 $\pm 0.0066$ with a
period of 137.8 $\pm 11$ years and a semi-major axis of 6.49 $\pm 0.35 \times 10^{16}$ cm.
The left panel of Figure 1 shows the orbit of the cloud. For
a Keplerian orbit, G2 would have been at apocenter
in the year 1944.6, with a distance from the SMBH of r$_{apo}$ = 1.26 $\pm 0.06 \times 10^{17}$ cm.
In 2011.5 its distance was 1.8 $\times 10^{16}$ cm. It will reach pericenter in 
2013.5 with a distance of merely r$_{peri}$ = 4.0 $\pm 0.3 \times 10^{15}$ cm. Due to its highly eccentric
orbit, the cloud spends most of its time at large radii r $\geq 5 \times 10^{16}$ cm. 

Our simulations (Section 6) show that the cloud will not
survive its pericenter passage. If it formed close to its presently
observed radius ("in situ" scenario, see Section 6.1), 
we would be very fortunate to have discovered it before it is
disrupted. Although such a coincidence cannot be ruled out,
it is more likely that the cloud formed close to apocenter at r $\geq 10^{17}$ cm at the inner
edge of the clockwise disk (shaded area in the left panel of Figure 1).

The right panel of Figure 1 shows the velocity of the cloud as function of its distance from the SMBH.
If G2 started at apocenter, its velocity was v=168 $\pm$ 80 km/s which is much smaller than its circular
velocity of 676 km/s. In 2011.5 its velocity had increased to v=2360 $\pm$ 50 km/s and at pericenter it
will have accelerated to a velocity of 5264 $\pm$ 300 km/s. For comparison, the dashed line
corresponds to the velocity expected for an object on a parabolic orbit.
The cloud is gravitationally bound to the SMBH with a binding energy per mass of

\begin{equation}
E_b/M_c = \frac{1}{2} v^2 - \frac{GM_{BH}}{r} = -4.4 \times 10^{15} \mathrm{erg/g} .
\end{equation}

\noindent For a total cloud mass of M$_c = 1.7 \times 10^{28} f_V$g, the total binding
energy would then be $E_b = -7.5 \times 10^{43} f_V$ erg. 

\section{INTERACTION WITH THE HOT SURROUNDING GAS}

Up to now we assumed a Keplerian orbit.
G2's motion is however affected by its hydrodynamical interactions with the confining,
hot gaseous surrounding (Gillessen et al. 2012). Here, we adopt the Yuan et al. (2003) model 
(see also Xu et al. 2006)  which reproduces the {\it Chandra} X-ray observations and is consistent with the low
accretion rate inferred by Marrone et al. (2007).
For a mean molecular weight $\mu = 0.614$ the
hot gas density distribution in the surrounding of Sgr A$^*$ is 

\begin{equation}
\rho_{hot}(r) = \rho_0 \ \frac{r_0}{r} = \eta_{hot} \times 9.5 \ 10^{-22}
\left(\frac{10^{16}\mathrm{cm}}{r} \right) \mathrm{g\cm}^{-3}
\end{equation}

\noindent where $r_0 = 10^{16}$ cm and $\rho_0 = \rho (r_0)$. 
$\eta_{hot} \leq 1$ takes into account the fact that 
some fraction of the observed  X-ray luminosity might be due to unresolved stellar sources
(Sazonov, Sunyaev \& Revnivtsev 2011). Given the density distribution, the assumption 
that the hot gas component is close to hydrostatic 
equilibrium within the SMBH's gravitational potential would lead to a temperature distribution 

\begin{equation}
T_{hot}(r) = T_0 \frac{r_0}{r} = \mu \frac{G M_{BH}}{2 R_g r_0} \ \frac{r_0}{r} = 
2 \times 10^8 \left( \frac{10^{16}cm}{r} \right) K
\end{equation}

\noindent with T$_0 = $T$(r_0)$. Note that T$_0$ is independent of the mass loading, i.e.
$\rho_0$. Our estimate of T$_0$ is a factor of 2 smaller than the value
adopted by Gillessen et al. (2012), taken from the Xu et al. (2006) model  with
a temperature of T$_0 = 3.5 \times 10^8$ K. It is however more consistent with 
Yuan et al. (2003, their Figure 2). If the hot gas would actually fall inwards
as expected for an accretion flow, then
T$_0$ would be even lower. If the temperature would be as high as adopted by Xu et al. (2006),
the pressure force exceeds the gravitational force, resulting in
an expansion which is in contradiction with the gas infall scenario.
The blue dotted line in the right panel of Figure 1 shows the
sound speed of the hot bubble, adopting equation 3. The cloud's
velocity is sonic with respect to the hot gas in the outer regions and becomes slightly supersonic (Mach 1.5) further in. 
We have assumed hydrostatic pressure equilibrium with the gravitational
potential of the SMBH. The situation is certainly much more complex. Rotation, convection, thermal winds,
heating and cooling processes, thermal conduction and
magnetic fields are likely to affect the detailed structure of
the hot bubble surrounding Sgr A$^*$ (e.g. Johnson \& Quataert 2007; Hawley \& Balbus 2002).
Given the limited amount of data it is currently difficult to constrain models
that include these effects. 
In order to keep the theory simple and well determined, we therefore restrict ourselves in this paper
to the simple atmosphere, as given by the equations 2 and 3. Our results and predictions might serve
as a first approximation and a test for subsequent, more sophisticated models.

A detailed investigation of the cloud's evolution and hydrodynamical interaction with
the surrounding requires an estimate of its radius R$_c$ and density $\rho_c$. 
In this paper we will follow Gillessen et al. (2012) and assume that the cloud
is ionized by the strong UV field of the central cluster of massive stars. Its cooling timescale is
(Sutherland \& Dopita 1993; Burkert \& Lin 2000) $\tau_{cool}  = 3 k_B T_c/ \Lambda n_c$
which for  a cooling rate of $\Lambda \approx 3 \times 10^{-22}$  erg cm$^3$ s$^{-1}$
and for cloud densities of n$_c \geq 10^5$ cm$^{-3}$ leads to $\tau_{cool} \leq 10^5$ sec
which is much shorter than the orbital period. G2 should therefore
maintain its temperature of $T_c = 10^4$ K during most of its evolution, set by photo ionisation 
equilibrium. Because of its small mass density with respect to its Roche density, self-gravity is completely negligible.
In this case, it will
try to achieve a homogeneous density state, in pressure equilibrium with the surrounding.
The hydrodynamical simulations, presented in Section 6 actually show a more complex structure
due to the compression of the front edge by ram pressure while at the same time 
the back, downstream tail is at slightly lower densities due to tidal shearing. 
This does however not affect much the cloud's effective size which, for most of its orbit,
is well reproduced by adopting a mean
density that is consistent with the assumption of pressure equilibrium:
$\rho_c = \rho_{hot} T_{hot}/T_c$.

Unfortunately, the observations do not resolve G2's minor axis, perpendicular to the orbital motion.
They only provide an upper limit of 12 mas, corresponding to $R_c \leq 1.5 \times 10^{15}$ cm. 
The 2008.3 position versus line-of-sight velocity (PV) diagram (left panel of Figure 5) shows
a velocity gradient that can be used in order to determine the elongation of the cloud
along its orbit which in this stage corresponds to a major axis radius of 21 mas or $2.6 \times 10^{15}$ cm
and a minor axis of $\leq 1.5 \times 10^{15}$ cm. Adopting a cylindrical shape as expected for a tidally elongated object,
a lower limit of its density in 2008 is then $\rho_c \geq 5 \times 10^{-19}$ g cm$^{-3}$. The cloud has been in this compact
state from the earliest spectral measurements that date back to 2002 till 2008.
In 2011.5 (middle panel of Figure 5) the cloud has developed a much stronger velocity
shear due to the gravitational acceleration. Its corresponding major-axis length is 22 mas or
2.75$ \times 10^{15}$cm, leading to a cloud density that has not changed much till
2008 if one adopts a constant minor axis radius.
As the minor-axis is not resolved, the cloud's density in 2011 could however
be substantially larger than in 2008 as it might have been stretched and compressed 
into a dense, thin, spaghetti-like filament (see Section 6).

Several non-linear and complex hydrodynamical processes dominate
G2's interaction with its surrounding, diffuse gas.
These include evaporation, ram pressure, gas stripping, cloud
compression or expansion as well as Rayleigh-Taylor and Kelvin-Helmholtz instabilities.
In addition, the cloud might be affected by compressional heating and cooling and might
be threaded by magnetic fields.
A detailed investigation of all these simultaneously acting processes 
requires self-consistent numerical simulations which are beyond the scope of
this paper. Here we restrict ourselves to some of the most basic aspects, leaving
more detailed models to subsequent papers.  We start with a 
discussion of the most dominant processes in isolation and
then present a first set of numerical simulations that investigate
especially the tidal effects onto G2's evolution.

\subsection{Cloud Evaporation}

Small gas clouds, embedded in a hot environment, will loose gas due to 
evaporation as a result of thermal conduction. The classical thermal conductivity
(Spitzer 1962; Parker 1963) is based on the diffusion approximation which assumes
that the electron mean free path $\lambda_e$ of the hot medium is small compared to
the cloud radius $R_c$. Cowie \& McKee (1977) in their seminal paper demonstrate
that this approximation is invalid as soon as $\lambda_e \geq R_c$ that is if the
saturation parameter $\sigma_0 = 1.84 \ \lambda_e/R_c > 1$ . In this case,
the heat flux is no longer determined by the classical diffusion formula.

For the central hot bubble we find

\begin{equation}
\lambda_e = 10^4 \left( \frac{T}{\mathrm{K}} \right)^2 \left( \frac{10^{-21} \mathrm{g cm}^{-3}}{\rho_{hot}} \right) 
\mathrm{cm} =
4 \times 10^{17} \left( \frac{10^{16}\mathrm{cm}}{r} \right) \mathrm{cm}
\end{equation}

\noindent which for all relevant distances r from the SMBH is indeed much larger than G2's observed 
size $R_c$. Adopting a typical cloud radius of order $1-2 \times 10^{15}$ cm the corresponding saturation 
parameter is $\sigma_0 \approx 50-500$. In the saturated limit and
assuming pressure equilibrium between the cloud and the hot surrounding,
the evaporation timescale is given by (Cowie \& McKee 1977)

\begin{equation}
\tau_{evap} = 120 yr \left( \frac{r}{10^{16}\mathrm{cm}} \right)^{1/6} 
\left( \frac{M_c}{1.7 \times 10^{28} \mathrm{g}} \right)^{1/3} .
\end{equation}

\noindent Cowie \& McKee (1977) argue that 
for a wide variety of conditions, the presence of a magnetic field does not strongly affect this timescale.
Interestingly, this evaporation timescale is very similar to the orbital period of 138 yr.
Equation 5 provides a strong lower limit on the mass of cold dust clouds that can penetrate as 
deeply into the hot bubble as observed for G2. 
Only clouds with masses of order or larger than G2 would reach distances of
1-2 $\times 10^{16}$ cm. We also can conclude that, once G2 breaks up at pericenter,
its subfragments might dissolve quickly. Note however that the mixing of cloud material
with the diffuse surrounding will change the thermodynamical properties of the bubble.
According to equation 2 the total mass of hot gas in the inner
$2 \times 10^{16}$ cm is similar to G2's mass. Mixing of cloud debris with the environment could
effectively cool the innermost region of the hot bubble which would reduce the efficiency of
evaporation. In addition, this process might destabilize the atmosphere which then collapses
onto the SMBH. The situation is different in the outer orbital parts where the mass of the hot bubble
is more than a factor of 100 larger that G2's mass.

\subsection{Ram pressure effects}

Ram pressure will compress the cloud into a lenticular and sickle-shaped 
structure (e.g. Murray \& Lin 2004; Schartmann, Krause \& Burkert 2011). The importance of ram pressure, compared
to the thermal pressure of the surrounding is given by the ratio v$^2$/c$_{hot}^2$, where c$_{hot}$
is the sound speed of the hot gas component. As shown by the blue dashed line in the right panel of Figure 1,
close to apocenter, gas pressure dominates while at its present location
ram pressure is a factor of 2-3 larger than the thermal pressure.

The drag will also remove part of the cloud's orbital energy.
Given the highly eccentric orbit, we can to a good approximation neglect the cloud's small
tangential motion. In this case, the energy lossed by G2 on its way from radius $r_1$
to radius $r_2$ is (Murray \& Lin 2004)

\begin{equation}
\Delta E_{drag} = \frac{1}{2} \pi C_D \int_{r_{1}}^{r_{2}} R_c^2 \rho_{hot}(r) v^2(r) dr .
\end{equation}

\noindent where $C_D \approx 1$ is the drag force. The observations already demonstrate that, within the observational
errors, G2 follows a Keplerian orbit. $\Delta E_{ram}$ therefore should be small, compared 
to the cloud's kinetic and absolute potential energy. In this case we can use $v(r)$ from
equation 1 and assume $E_b/M_c$ to be roughly constant. For a constant cloud radius $R_c$
the integration of equation 6 leads to 

\begin{equation}
\Delta E_{ram}(r) = \pi \rho_0 r_0 C_D R_c^2 \left[ \frac{E_b}{M_c} \ln \left(\frac{r_2}{r_1} \right)
- G M_{BH} \left(\frac{1}{r_2} - \frac{1}{r_{1}} \right) \right] .
\end{equation}

\noindent Adopting $C_D = 1$ and an effective cloud radius of $R_c=2 \times 10^{15}$ cm,
the energy lossed between apocenter and
$r_{2011.5}=1.8 \times 10^{16}$cm due to ram pressure effects would be
$\Delta E_{ram}(r_{2011.5})= 2 \eta_{hot} \times 10^{42} $erg = $ 0.03 \eta_{hot} E_b$.
This confirms that the cloud moves on a ballistic orbit during
its approach of the SMBH. It also indicates that our estimate of G2's apocenter
is robust. For such an extended cloud, ram pressure effects would however
change the cloud's orbit substantially within the next
2 years on its way towards pericenter with 
$\Delta E_{ram}(r_{peri})= 2  \eta_{hot} \times 10^{43}$ erg = 0.3 $\eta_{hot} E_b$. 
It is unlikely to assume a fixed size as the cloud will continuously try to achieve
pressure equilibrium with its surrounding. If $\rho_c =\rho_{hot} T_{hot}/T_c$ and
adopting a spherical geometry the energy loss is

\begin{equation}
\Delta E_{ram}(r) = \pi \eta_{hot}^{1/3}\left(\frac{\rho_0}{r_0}\right)^{1/3} 
\left(\frac{3 M_c T_c}{4 \pi T_0} \right)^{2/3} \left[ \frac{3}{4}\frac{E_b}{M_c} 
\left(r_2^{4/3}-r_1^{4/3}\right) + 3G M_{BH} \left(r_2^{1/3}-r_1^{1/3}\right) \right].
\end{equation}

\noindent Now, for $T_c=10^4$ K, on its way from apocenter to $r_{2011.5}$, the cloud would loose
$\Delta E_{ram}(2011.5)= 0.02 \eta_{hot}^{1/3}  E_b$. This value increases to 0.04 $\eta_{hot}^{1/3}$
till pericenter which is much smaller than the constant radius case due to the strong compression of the 
cloud at small orbital radii. To a good approximation, the cloud should then
stay on a Keplerian orbit till pericenter.

\subsection{Pressure Confinement, Implosion and Hydrodynamical Instabilities}

As self-gravity is negligible, for a constant temperature $T_c$, the cloud will 
try to achieve a homogeneous density state, in thermal pressure equilibrium with the surrounding.
The shaded region in Figure 2 indicates the expected cloud density $\rho_{equi}$ for the case 
of pressure equilibrium with the surrounding.
The thick black line shows $\rho_{equi}$, adopting
a hot gas temperature and density as given by the equations 2 and 3, with $T_c = 10^4$ K. 
The lower and upper boundaries correspond to $\eta_{hot}=0.5$ and $T_0 = 2 \times 10^8$ K
and $\eta_{hot}=1$, $T_0 = 3.5 \times 10^8$ K, respectively.
Labeled black points show G2's estimated density assuming its
minor axis is marginally resolved. In this case, the cloud would have been in pressure equilibrium
in 2003. $\rho_{equi}$ increases strongly between
2003 and 2011. In contrast, the observations reveal an increasing velocity gradient
in the cloud as expected with decreasing orbital radius. Interestingly, the change in the
velocity gradient is consistent with the assumption that G2's major axis remains roughly
constant. The assumption of a minor axis close to the resolution limit then leads to a constant 
cloud density after 2003 that does not change significantly with time. The
cloud's thickness in the unresolved direction perpendicular to its motion might however decrease
continuously.  In this case $\rho_c$ would increase with time.
The 2008 and 2011  points in Figure 2 then represent lower limits.

We unfortunately do not have detailed information about the cloud's structure prior to 2003. One can however
investigate the question whether a cloud like G2 could have achieved pressure equilibrium in 2003 if
it started with an arbitrary density at apocenter. The implosion of a gas clump
that is imposed to a high pressure environment has been studied in details by Klein, McKee \&
Colella (1994).  A shock front forms at the outer edge that sweeps up the cloud.
It accelerates till the sum of the ram-pressure of the shock
$\rho_c v_s^2$ and the internal gas pressure $P_c = \rho_c c_c^2$ with c$_c$ the cloud's internal sound speed
is equal to the external pressure $P_{hot}=\rho_{hot} c_{hot}^2$, with c$_{hot}$
the local sound speed of the surrounding gas, leading to an equilibrium shock velocity of

\begin{equation}
v_s^2 = \left( \frac{\rho_{hot}}{\rho_c} \right) c_{hot}^2 - c_c^2.
\end{equation}

\noindent The cloud's crushing timescale is then

\begin{equation}
\tau_{cc} = \frac{R_c}{v_s} = 0.5 \tau_{s} \left( \frac{P_{hot}}{P_c}-1 \right)^{-1/2}
\end{equation}

\noindent where $\tau_{s}=2 R_c/c_c$ is the cloud's sound crossing timescale which for our typical
parameters of a $10^4$~K gas clump, in an environment, 
described by the equations 2 and 3 leads to

\begin{equation}
\tau_{s} = 2 \left(\frac{3 M_c}{4 \pi c_c P_{c}} \right)^{1/3} = 176 \ \eta_{hot}^{-1/3}
\left(\frac{r}{r_{apo}}\right)^{2/3} \left(\frac{P_{hot}}{P_c} \right)^{1/3} yr .
\end{equation}

\noindent Figure 3 shows G2's sound crossing timescale (red shaded region) as function of orbital radius,
assuming pressure equilibrium $P_{hot}=P_c$. Interestingly, close to apocenter, $\tau_{s}$ is almost equal to
the dynamical time (orbital period) shown by the horizontal dotted line.

Equation 10 demonstrates that for $P_{hot} >> P_c$ the crushing timescale of the cloud
is much smaller than its sound crossing time and dynamical timescale. Starting close to apocenter,
it should therefore implode long before reaching its current position. 
The dashed, red lines in Figure 2 show the evolution of the mean density of such a cloud.
For that, we integrated the evolution of the compression front, taking into account the time variation of 
$P_{hot}(r)$ as the cloud moves through the stratified environment with the pressure
increasing with decreasing distance r(t) from the SMBH. 
Assuming a constant cloud temperature, the evolution of the outer cloud radius is then given by
$dR_c/dt=-v_s(t)$ where $v_s(t)$ depends on the orbital radius r(t) according to equation 9. 
Note that these clouds, when
crossing the shaded area of pressure equilibrium, continue to contract.
Clouds that are in pressure equilibrium are one-component systems with
a constant gas density. In contrast, the imploding clouds, despite the fact
that their volume averaged density is the same, contain an inner, low-pressure and low-density
core that is being swept up by a high-density, inwards moving shell. 

As the high-density shock front is accelerated by a low-density gaseous environment the cloud
will be shredded by Rayleigh-Taylor instabilities that grow on a timescale of (Klein, McKee \& 
Colella 1994)

\begin{equation}
\tau_{RT} = \frac{\tau_{cc}}{(k R_c)^{1/2}} 
\end{equation}

\noindent The shortest wavelength or largest wave number $k$ has the fastest growth rate. But
Rayleigh- Taylor perturbations saturate in the non-linear regime. As a result, the most
destructive wavelength is given by $k R_c \approx 1$. The cloud should therefore be destroyed
and mix with the surrounding on a timescale of order its crushing timescale. In summary, under-pressured
clouds, starting at apocenter, would disappear before reaching the presently observed position.
If G2 came from apocenter, it therefore was either in pressure equilibrium with the surrounding
or had an even larger pressure. 

Even if G2 is initially in pressure equilibrium, the pressure of the surrounding
hot gas increases fast on its way towards the SMBH. The cloud will be able to adjust
as long as $\tau_{s}$ is short compared to the timescale $\tau_p$ 
on which the external pressure changes 

\begin{equation}
\tau_p = \frac{1}{v_r d \ln P_{hot}/dr} = \frac{r}{2 v_r}
\end{equation}

\noindent with v$_r$ the radial velocity of the cloud. The solid black line in Figure 3 shows
$\tau_p (r)$ which for G2's orbit becomes smaller than $\tau_{s}$ quickly for r $\leq 10^{17}$ cm.
Entering this region, the cloud will become a cold, low-pressure island within a high-pressure 
surrounding and start imploding. 
The evolution of such a system is shown by the thick, blue
line in Figure 2 where we have assumed
that the implosion starts as soon as $\tau_p < \tau_{s}$. In 2003, G2's mean density should still be 
close to $\rho_{equi}$. At that point in the evolution however, 
the timescale for the shock to completely crunch the cloud has
become smaller than the local infall timescale $\tau_{infall}$ and its density now increases fast.
Here we have neglected tidal shearing. As we will show
in Section 6 tidal effects actually reverse the evolution, leading
to a decreasing density when the cloud approaches the inner regions of its orbit.

Within the framework of the "in situ" scenario, discussed in Chapter 2, the cloud might have formed
half-way between apocenter and the SMBH, e.g. in the year 1995, shortly before G2 was detected. 
The thick, black dashed line in Figure 2 shows the clump implosion for this case.
Due to G2's relatively high initial density,
the velocity v$_s$ of the inwards moving shock is now less fast and will not compress
the cloud significantly within the next 12 years while the pressure in the surrounding rises steeply.
The cloud therefore evolves isochorically, in good agreement with the observations.

G2 might have started at apocenter in a high-density state with a pressure that was
larger than in the surrounding gas. In this case, the cloud will first expand with a velocity that
should be roughly equal to its sound velocity, trying to achieve pressure
equilibrium with the environment. This expansion phase is shown by the solid red line in Figure 2.
The subsequent evolution is complex and requires numerical simulations that will be discussed in Section 6.

\section{Probing the Temperature and Density Structure of the Central Hot Bubble}

G2 is a sensitive probe of the conditions in the central hot bubble.
Especially the early observations around 2002-2004, when the cloud is at a distance
of $r\approx 5.5 \times 10^{16}$ cm provide important constraints on the density 
and pressure of the surrounding diffuse
gas. In the following we will adopt the radial dependence of $\rho_{hot}$ and T$_{hot}$
as given by the equations 2 and 3. In addition,
we assume that the cloud is in pressure equilibrium with its surrounding in these early phases.
As discussed in Section 4, a lower limit of the density of G2 in 2002-2008 
is $\rho_c=5 \times 10^{-19}$ g cm$^{-3}$. For a cloud temperature of T$_c=10^4$ K this implies 
$\rho_{hot,2002} T_{hot,2002} = 0.033 \times \rho_0 T_0 \geq 5 \times 10^{-15}$ K \ g/cm$^{3}$
or

\begin{equation}
\rho_0 \times T_0 \geq 1.5 \times 10^{-13} \mathrm{g cm^{-3} K}.
\end{equation}

\noindent If $\rho_0 \leq \eta_{hot} \times 10^{-21}$ g cm$^{-3}$,
as inferred from the {\it Chandra} observations (Xu et al. 2006),
$T_0 \geq 1.5 \times 10^8 / \eta_{hot}$ K which is in agreement with Equation 3
if $\eta_{hot} = 0.75$. This would indicate that the fraction $\eta_{hot}$ 
of X-ray luminosity, resulting
from unresolved stellar sources (Sazonov, Sunyaev \& Revnivtsev 2011), is small as
$T_0$ cannot be much larger than a few $2 \times 10^8$ K if the hot bubble is bound to the SMBH.

\section{Hydrodynamical Simulations}
\label{sec:numhd}

In order to validate the results obtained from our analytical estimates
and to investigate the future evolution of the cloud
we have conducted idealised hydrodynamical simulations.

The hydrodynamical equations were integrated with the help of 
{\sc PLUTO}, version 3.1.1 (Mignone et al. 2007) which is 
a fully MPI-parallelized, high resolution shock capturing scheme
with a large variety of Riemann-solvers. 
For all simulations shown in this article, we propagated the state vector with the two-shock Riemann solver,  
did a parabolic interpolation and employed the third order Runge-Kutta time integration scheme.
In these first simulations, we were interested in the evolution of the cloud in an idealised 
atmosphere with a smooth density and pressure distribution in concordance with observations, as
discussed in Section 4. In order not to be dominated by the interaction of the
cloud with disturbances of the convectively unstable atmosphere, we artificially stabilise it.
This was done by additionally evolving a passive tracer field ($0 \leq tr \leq 1$), which allowed us to distinguish 
between those parts of the atmosphere which have interacted with the cloud ($tr \geq 10^{-4}$) from those which changed
due to the atmosphere's inherent instability ($tr \leq 10^{-4}$). Those cells fulfilling the latter criterion 
were reset to the values expected in hydrostatic equilibrium. The boundary conditions were set to the values expected
for hydrostatic equilibrium, enabling outflow but no inflow.
The adiabatic index $\Gamma$ was set to one, which we consider 
a reasonable assumption, as the temperature structure of the advection-dominated accretion
flow solutions (e.g. Narayan \& Yi 1994; Narayan 2002) is supposed to
be given by adiabatic heating of the accretion flow itself
and the temperature of the cloud material is expected to be set by photoionisation equilibrium
in the radiation field of the surrounding stars (Gillessen et al. 2012).
Our two-dimensional computational domain represents the orbital plane of the cloud, which initially starts in pressure 
equilibrium with the atmosphere except for simulation CC03, initially having an overpressure of a factor of 100. 
The corresponding parameters are summarised in Table 1. The cloud starts on the negative x-axis
of the fixed cartesian coordinate system with a spatial resolution of $7\times 10^{13}$cm ranging from  
$-1.3\times 10^{17}$cm to $1.2\times 10^{16}$cm in x-direction
and $-6.2\times 10^{16}$cm to $2.5\times 10^{16}$cm in y-direction.
It orbits in clockwise direction with the major axis parallel to the x-axis and the pericenter 
of the orbit on the positive x-axis. The black hole is located at the origin of our coordinate system.
We neglect magnetic fields as well as feedback from the central source for the sake of simplicity 
and will give a more detailed analysis of the simulations and numerical tests of this approach 
in Schartmann et al. (2012).

\subsection{Model CC01: In Situ Formation of G2}

Figure 4 shows the evolution of a cloud that was born as a spherical droplet in the year 1995
at a distance of 7.6 $\times 10^{16}$ cm. Dotted white
contours depict the expected shape of the cloud from a collisionless test particle simulation, 
if each gas particle would
move on a ballistic orbit within the gravitational potential of the SMBH. In 2008.5
the hydrodynamical interaction with the surrounding is still small and the cloud's structure 
follows the outer contour of the test particle
simulations very well. The cloud already in this early phase is developing an elongated structure
due to tidal effects. Ram pressure stripping at the outer edge also generates
a trailing tail of gas that is dispersed by Kelvin-Helmholtz instabilities. The mass loss is however negligible.
In 2011.5 and 2012.5 the cloud has become significantly more elongated.  Ram pressure effects are now
more clearly visible at the front side which is falling behind the ballistic orbits.
The dotted black line in Figure 2 shows the mean density evolution as function of orbital radius.
In agreement with our analytical estimates (dashed black line in Figure 2), the cloud's density
is not changing significantly between 1995 and 2011 as the crushing timescale for an object
that starts in 1995 is longer than its infall timescale. While the mean density in the idealized model is
continuously rising due to compression, it actually decreases slowly in the numerical simulation
due to the tidal shearing along the orbit. This might be partly an artifact of the 2-dimensional simulations
and needs to be confirmed by future 3-dimensional models.

Figure 5 compares the structure of the simulated cloud in the 
PV-diagram with the observations. 
In the year 2008.5 the cloud is still compact. In 2011.5 it has 
developed a strong shear that follows the center-of-mass, ballistic orbit (dashed line). Overall 
the structure of the simulated cloud in the PV-diagram is in excellent agreement with the 
observations (contours).
The predicted structure and kinematics in 2012.5 is shown in the 3rd panels of the Figures 4 and 5
when the cloud has reached the point of maximum line-of-sight velocity. In the PV-diagram
the cloud's contours still do not deviate significantly from a ballistic orbit, shown by the dotted lines.

Figure 6 shows G2's structure for the "in situ" scenario during and after pericenter passage.
At pericenter the cloud will be tidally stretched into a long, curved filament. 
As discussed in Section 4.2 the hydrodynamical drag is not strong enough to remove
a large fraction of the cloud's kinetic energy. As a result, the bulk of the gas is
moving out to large distances again. Ram pressure
at the head has however removed a substantial amount of kinetic energy and angular momentum forcing
the front of the filament to fall into the unresolved inner accretion zone of Sgr A$^*$
(upper panel). The middle panel of Figure 6 shows that on its way outwards 
the gas is first compressed both by the negative divergence of the velocity field and by ram pressure.
Large Kelvin-Helmholtz eddies are now visible at the inner edge, facing the SMBH, that
begin to destroy the cloud, leading to several narrow gas streams that fall into Sgr A$^*$.
In 2050.5 the cloud has completely dissolved and generated a narrow, dense stream of cold gas that
feeds the SMBH.

\subsection{Model CC02: Formation of G2 at Apocenter}

We have argued in the Sections 2 and 3 that it is more likely for the cloud to have started in the clockwise
rotating stellar disk which agrees with the cloud's orbital plane and which has an inner
edge equal to its apocenter. Figure 7 shows the evolution of an initially
spherical, cold gas droplet in pressure equilibrium at apocenter. Initially the cloud remains
in pressure equilibrium with its surrounding,
in agreement with the analytical estimate (blue line in Figure 2).  However even in the early
phases tidal effects begin to reshape the cloud. The upper panel of Figure 8 shows the characteristic
velocity field in the cloud at 1980.6 after subtracting its center-of-mass motion (large black arrow).
The mean cloud density is $2 \times 10^{-19}$ g cm$^{-3}$, in agreement with our analytical estimates.
However, despite the fact that the cloud is still at a distance of $10^{17}$ cm,
the strong tidal shear has already generated a strong internal velocity field that
pushes material outwards along the orbit while at the same time compressing the 
cloud perpendicular to it. The surrounding gas pressure, even in combination with the ram pressure
is not strong enough in order to stop this flow. Note also that the cloud has 
developed a density gradient due to ram pressure compression at the front. The lower panel of Figure 8
shows the cloud at 2003.6 when it is at a distance of $5.3 \times 10^{16}$ cm.
Its mean density has increased to $\rho_c = 10^{-18}$ g cm$^{-3}$ which is in agreement
with the observationally  inferred cloud density (Figure 2). It is however in disagreement
with our analytical estimates that predict higher compression if tidal effects are neglected. 
This is due to the fact that at
this stage tides have turned the cloud into a long filament. The situation has become even more extreme in 2011.6
(middle panel of Figure 7) where the filament has grown to a length of 
$3 \times 10^{16}$ cm which is much longer than the observationally inferred length of $\leq 3 \times 10^{15}$ cm.
Note that the filament is distorted by
Kelvin-Helmholz instabilities as a result of its interaction with the diffuse surrounding. 
In 2012.6 (lower panel of Figure 7), these
instabilites have become non-linear and begin to break up the system into a string of clumps
that begin to fall into the SMBH. The PV-diagramm shown in Figure 9 confirms that the structure is much too
elongated compared to the observations.

\subsection{Model CC03: An Overpressured High-Density Cloud at Apocenter}
We argued in the analytical part (Section 4) that a cloud that has a pressure
much larger than its surrounding will expand. The detailed evolution is
however difficult to predict analytically. We therefore show in Figure 10 the evolution
of a clump that started at apocenter with an initial density that is a factor of
100 larger than the equilibrium density $\rho_{equi}$. The cloud initially expands, but the
expansion does not stop as soon as the mean cloud density has become
equal to $\rho_{equi}$ due to the outwards directed velocity field. The cloud
overshoots this equilibrium point and its density and pressure drop further
with the cloud forming a very extended structure that is now subject to
the strong tidal forces as shown in Figure 10. Like in the previous case, the
PV diagramm (Figure 11) is not at all in agreement with the observations.
This confirms our previous conclusion that, if G2 is a diffuse cloud, it started
its journey into the center close to pressure equilibrium with the surrounding.

\subsection{Model SS01: The Spherical Shell Model}

Up to now we assumed that G2 is an isolated, small few Earth mass cloud that formed
either close to the location of first detection or at apocenter. It might however
be part of a larger structure. The observations (Figure 2 of Gillessen et al. 2012; 
contours in the PV diagrams, e.g. Figure 5)
indeed show extended emission downstream of G2 with a brighter area 
at the end that is offset with respect to G2's orbit with a line-of-sight velocity difference of order 500 km/s.
We have tried to generate an initial condition at apocenter that could reproduce these
observations and found that a surprisingly simple symmetric shell of gas in the orbital plane
with a tangential velocity of 125 km/s, similar to G2's initial velocity,
an outer radius of $1.2 \times 10^{16}$ cm and a thickness of $3 \times 10^{15}$ cm can 
reproduce the observations well (table 1). Its gas density $\rho_c = 1.42 \times 10^{-19}$ g cm$^{-3}$ 
is given by the requirement that it is in pressure equilibrium with its surrounding. Such a structure could have been
produced either by the dusty wind shells of massive stars, an explosion or
an object that crossed the plane of the stellar disk at apocenter. Figure 12 shows the 
evolution of the shell. G2 in this case represents its leading head that is being compressed
into a spheroidal clump as a result of the converging, radial inflow. Interestingly, the
projection of this distorted shell along the line of sight also generates a second
high surface density region that is connected to G2 by low surface density gas.
This is clearly shown in the shell's PV diagram (Figure 13). Note that this second bright
point is also offset with respect to G2's orbit (dashed line) by $\sim$ 500  km/s,
as observed. 

\section{DISCUSSION AND CONCLUSIONS}

We investigated the origin and evolution of the small gas cloud G2 that has been
detected approaching the Galactic SMBH on a highly eccentric orbit. The cloud
has an observationally inferred density and temperature that during the first
years of detection around 2003 indicates that it was in pressure equilibrium with
its surrounding. Our analytical estimates and numerical simulations show
that it must have formed close to pressure equilibrium. A cloud with a
significant lower pressure would be destroyed by implosion while a highly overpressured
cloud would go through a phase of rapid expansion, forming a diffuse, extended
and tidally elongated structure that is not in agreement with the observations.
The "in situ" cloud scenario which assumes that the cloud formed as a spheroidal
clump in pressure equilibrium in the year 1995, shortly before it was detected,
provides an excellent match to all the observations, dating back to 2002 till today.
In this case, our numerical simulations indicate that the cloud will be tidally stretched into
a very elongated filament till 2013.5 which shortly after pericenter passage will begin to 
loose gas as a result of Kelvin-Helmholtz instabilities. This gas might fall into the
inner accretion zone of Sgr A$^*$, forming a hot luminous accretion disk.
Over the next 80 years the debris of the tidally destroyed cloud will
fuel the SMBH, maybe leading to an extended period of nuclear activity. 

The main caveat of the "in situ" model is the fact that we have not been able
to identify any physical mechanism that could have formed G2 in 1995. We did not
find any known star that was close to its birth place at that time. In addition,
the  cloud could also not have formed by a cooling instability of the hot gas
as its cooling timescale is much longer than the dynamical timescale (Burkert \& Lin 2000).

It is intriguing that the assumption of G2 having formed at apocenter leads
to several interesting correlations that should be reproduced by any theoretical
model of its formation. G2's orbital plane coincides with the plane of the clockwise
rotating stellar disk and its apocenter agrees with the inner disk edge.
In addition, G2's sound crossing timescale, evaporation timescale
and orbital timescale are all equal at apocenter.
If this is not a coincidence it provides several important
constraints on its formation. First of all, G2 obviously did not form very
recently (e.g. 1995) by some condensation processes but came from apocenter. It also means
that G2 has been in pressure equilibrium at apocenter, either as a result of its
formation or due to re-adjustment. It is reasonable to assume that a cloud
that is subject to distortions when interacting with a turbulent environment
can only maintain coherence if it can react and re-adjust to them. This requires that
the cloud's sound crossing timescale, which is the timescale on which information
is transported through the cloud, is smaller than the timescale of the perturbation.
Clouds that cannot react will break up into subunits that are smaller
than this characteristic timescale. The origin of these perturbation is not clear up to now.
The inner hot bubble is probably highly turbulent and might even be convective. 
For a Kolmogoroff spectrum, the largest fluctuations which are of order the size of the 
region and which interact with the cloud on timescales of order the dynamical time
would dominate and might regulate its size. In this case,
G2 might represent a piece of a larger dusty cold region that disintegrated into
smaller clumps by its interaction with the turbulent environment. As the timescale
for such a region to break up is at least equal to an orbital period, 
the cold gas inside G2 must be older than that. G2
should then have completed at least one full orbit. The massive progenitor cloud might
already have been on a highly eccentric orbit at that time. Another possibility is that
it  was on a circular orbit.  within the stellar disk region, where it experienced a collision or violent 
interaction e.g. with a stellar wind shell. This destroyed the cloud
while at the same time deflected some of its parts onto highly eccentric orbits like G2.

One of the problems with all formation scenarios that start at apocenter  is the fact that
G2 will become very elongated with a strong
velocity gradient around 2011 that is not in agreement with the observations. As our simulations show, 
one possible solution is that G2 is actually part of a larger shell. This scenario could explain the
observed extended tail of lower surface brightness Br$\gamma$ emission that is seen on about the same
orbit as that of G2. The shell's origin is not clear. It might represent a clumpy wind shell,
ejected from one of the luminous blue variable or Wolf-Rayet stars in the stellar disk. 
We investigated which of the known
high-mass stars in the stellar disk was
close to G2's apocenter in the year 1944 and found one candidate, S91 which
appears to be an O6.5II star with a mass loss rate of order $1 \pm 0.5 \times 10^6$ M$_{\odot}$/yr and
a wind velocity of 2000 $\pm 500$ km/s (Puls \& Kudritzki, private information).
Winds of this kind are usually believed to generate hot plasma with temperatures of order  $10^8$ K
(Hall, Kleinmann \& Scoville 1982; Cuadra et al. 2005). Whether a cooling instability in the expanding
wind bubble could also have generated a ring of cold gas is not clear yet.

If the diffuse cloud scenario of G2 fails we are left with the possibility that G2
actually is the visible diffuse gas atmosphere of an unresolved, dense object in its center
(Murray-Clay \& Loeb 2012). This would
keep the cloud spheroidal, despite the external gravitational force of the SMBH,
first of all because gas is continuously replenished from the probably spherical radial outflow
and secondly, because the gravitational force of the central object inside the Roche volume
can balance the destructive and deforming gravitational force of the SMBH. A crucial
test of this scenario is, whether it can explain the close
agreement between the sound crossing, evaporation and dynamical timescales, that we discussed earlier.
In this case, one also would expect a tail of stripped material, trailing G2. Its structure and
especially its velocity distribution
might however differ significantly from the shell scenario. For example, it is not clear whether its 
downstream end would be bright. In addition, it is unlikely that the trailing stripped gas will
have the observed offset towards large infall velocities compared to the orbital velocity. We would
instead expect smaller infall velocities due to the deceleration by the ram pressure which is not observed.
Detailed observations of the diffuse environment of G2 and additional numerical simulations
would be helpful in distinguishing between the diffuse cloud and compact source scenario.

We have presented and discussed a first set of simplified simulations, adopting a hydrostatic atmosphere
and an isothermal cloud. Already these simulations reveal a very complex, non-linear evolution.
Future papers should add the effects of compressional heating of the cloud, especially close
to pericenter where Gillessen et al. (2012) estimate compression to significantly increase the temperature
and by that the luminosity of the cloud.
First test simulations demonstrate however that this effect does not significantly alter the
cloud's evolution as the gravitational force of the SMBH close to pericenter is much stronger
than the increase in gas pressure due to tidal heating. After pericenter the cloud material will
actually cool again due to expansion. It is also
important to investigate the effects of gas evaporation and magnetic fields. In the long
run one should replace the currently hydrostatic atmosphere by a turbulent and probably convective, 
inhomogeneous and rotating hot gas bubble that might strongly affect the cloud evolution.

The problem of the origin and evolution of the tiny gas cloud G2, falling into the accretion zone
of Sgr A$^*$ represents an exciting challenge for numerical and theoretical models of the
complex multi-phase
gas physics in the Galactic Center. We are in a unique situation where theoretical models
and numerical simulations of G2's evolution will be tested directly
by observations within the next couple of years. For the next decades
G2's journey through the Galactic nucleus and its
evolution will provide detailed information about this fascinating and extreme environment
of the Milky Way and the processes that feed its central SMBH.

\begin{table}
\begin{center}
\caption{Parameters of the hydrodynamical simulations.\label{tab:simparam}}
\begin{tabular}{llllllll}
\tableline\tableline
  & $\tau_{0}$\tablenotemark{a} &  $\rho_{\mathrm{cloud}}$\tablenotemark{b} & $R_{\mathrm{cloud}}$\tablenotemark{c} &
$x_{\mathrm{ini}}$\tablenotemark{d} & $y_{\mathrm{ini}}$\tablenotemark{e} & $v^x_{\mathrm{ini}}$\tablenotemark{f} &
$v^y_{\mathrm{ini}}$\tablenotemark{g} \\
 & yr AD & $10^{-19}\,\mathrm{g}\,\mathrm{cm}^{-3}$ & $10^{15}\,\mathrm{cm}$ & $10^{16}\,\mathrm{cm}$ & $10^{16}\,\mathrm{cm}$ 
& $\mathrm{km}\,\mathrm{s}^{-1}$ & $\mathrm{km}\,\mathrm{s}^{-1}$ \\
\tableline
CC01 & $1995.5$ & $6.21$   & $1.87$ & $-7.22$  & $ 2.21$ & $794.59$ & $48.45$\\
CC02 & $1944.6$ & $2.24$   & $2.63$ & $-12.59$ & $0.0$   & $0.0$    & $167.29$\\
CC03 & $1944.6$ & $223.64$ & $0.57$ & $-12.59$ & $0.0$   & $0.0$    & $167.29$\\
SS01  & $1927.2$ & $1.42$ & $11.8$ & $-15.80$ & $0.0$ & $0.0$ & $125.00$ \\
\tableline
\end{tabular}
\tablenotetext{a}{Start time of the simulation.}
\tablenotetext{b}{Initial density of the cloud.}
\tablenotetext{c}{Initial radius of the cloud.}
\tablenotetext{d}{Initial x-position of the cloud.}
\tablenotetext{e}{Initial y-position of the cloud.}
\tablenotetext{f}{Initial x-velocity of the cloud.}
\tablenotetext{g}{Initial y-velocity of the cloud.}
\tablecomments{CC refers to simulations of the compact cloud scenario and SS to those of the spherical shell scenario.}
\end{center}
\end{table}

\begin{figure}[ht]
\begin{center}
\includegraphics[width=1.0\textwidth]{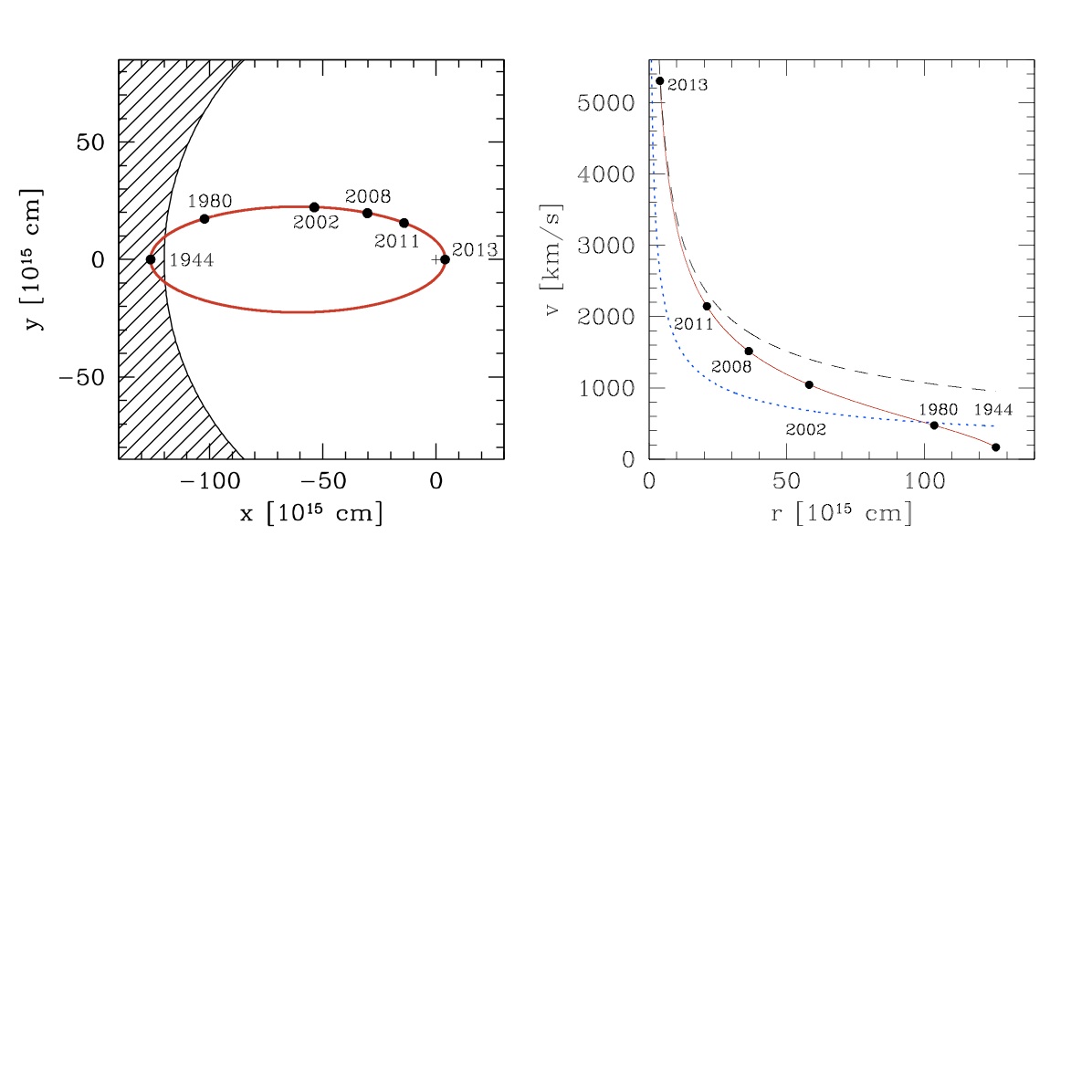}
\end{center}
\caption{
\label{fig:one} The left panel shows the orbit and location of G2 in the orbital plane. G2's apocenter
coincides with the inner edge of the clockwise rotating stellar disk (shaded area). The cloud will 
pass the SMBH (cross) in 2013 at a distance of merely $\sim 3100$ times the event horizon. Due to its highly
eccentric orbit, the cloud spends most of its time in the outer regions.
The right panel shows the velocity of G2 as function of its distance r from the SMBH. For comparison,
the dashed line shows a parabolic, unbound orbit with total energy E=0. The dotted line shows
the local sound speed of the diffuse, hot environment (Equation 3).}
\end{figure}

\begin{figure}[ht]
\begin{center}
\includegraphics[width=0.8\textwidth]{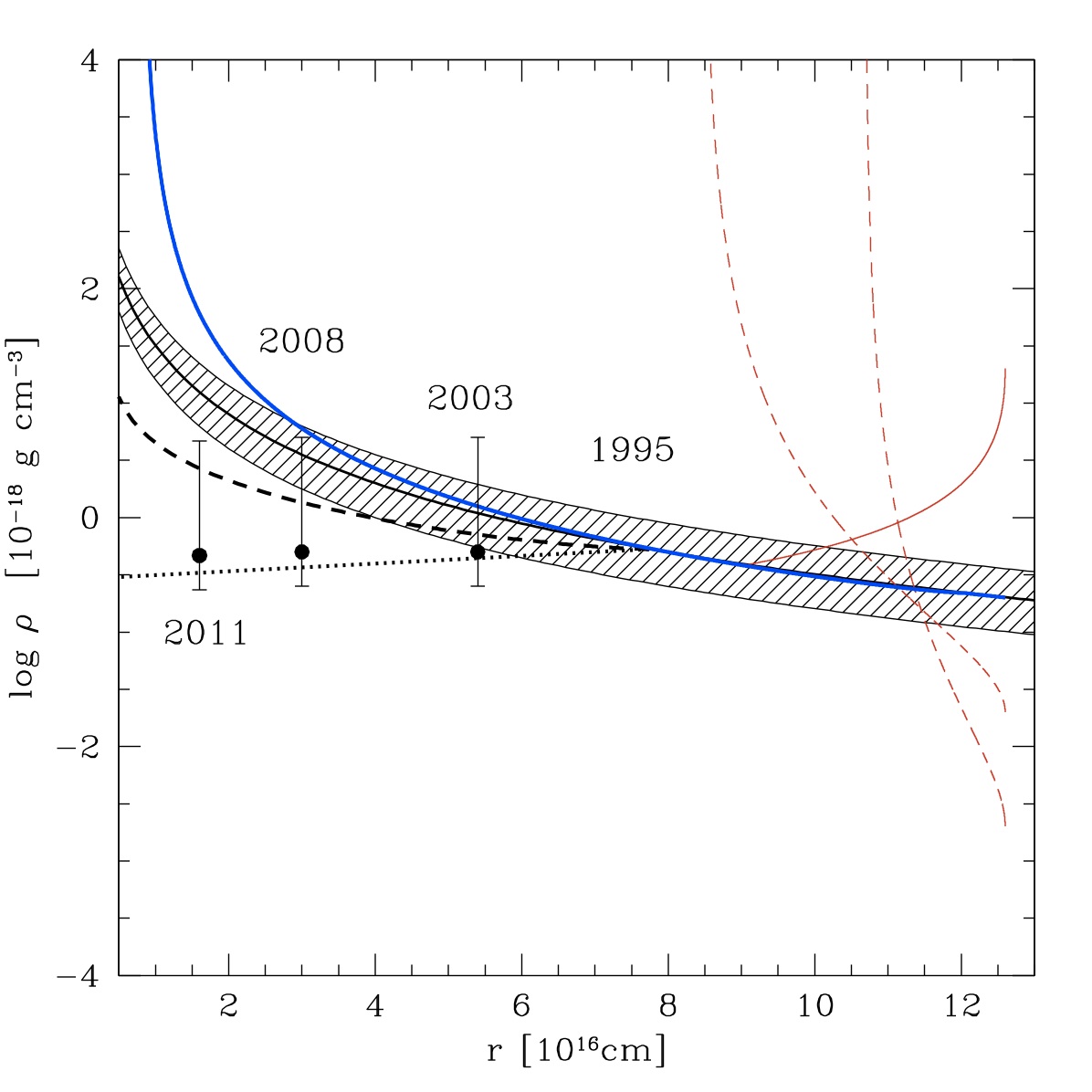}
\end{center}
\caption{The dashed region shows the expected density of G2 as function of orbital radius r, assuming pressure equilibrium
with the surrounding hot bubble. The three labeled data points show lower limits of the observationally 
inferred cloud density. All lines correspond to the evolution of a cloud in a time varying external gas component
with a pressure that increases with time as expected for G2 on its orbit into the center. 
With the exception of the dotted black line, tidal effects are neglected.
The thick dashed line
shows the theoretically expected density of G2 if it formed as a spherical cloud in the year 1995, neglecting tidal effects.
For comparison, the dotted black line 
shows the mean density evolution of the numerical simulation CC01, starting also in 1995 (Section 6.1), where tidal effects are included.
The thick blue line shows the density evolution if G2 started at apocenter. Dashed red lines correspond to the evolution
of a cloud that starts with a pressure that is a factor of 10 or 100 smaller at apocenter than required for pressure equilibrium.
The solid red line shows the early expansion phase of an overpressured cloud with initial density that is a factor of 100
larger than the equilibrium density. 
\label{fig:two} 
}
\end{figure}

\begin{figure}[ht]
\begin{center}
\includegraphics[width=1.0\textwidth]{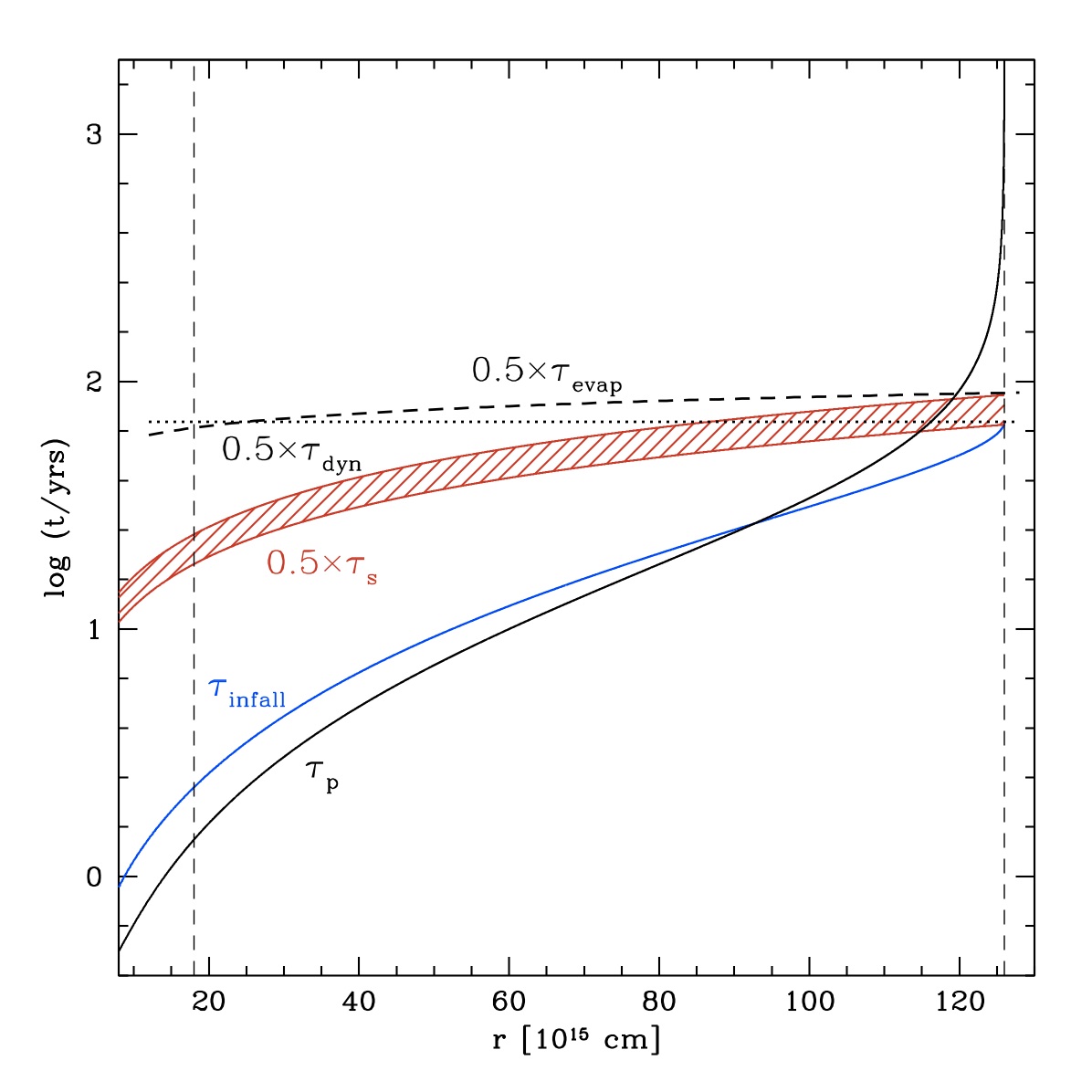}
\end{center}
\caption{The various timescales are shown as function of distance r from the SMBH
that affect G2's evolution: the dynamical timescale $\tau_{dyn}$ (black, horizontal dotted line), 
the evaporation timescale $\tau_{evap}$ (black dashed line), 
the sound crossing timescale $\tau_{s}$ (shaded red region), adopting $\eta_{hot}=1$ (upper limit)
and  $\eta_{hot}=0.75$ (lower limit), respectively, the pressure
timescale $\tau_{p}$ (black solid line) and the infall timescale $\tau_{infall}$ (blue solid line)
that measures the time it takes G2 to reach pericenter from radius r. The two vertical dashed lines
correspond to the apocenter and the location of G2 in 2011.3, respectively.
\label{fig:three} }
\end{figure}

\begin{figure}[ht]
\begin{center}
\includegraphics[width=1.0\textwidth]{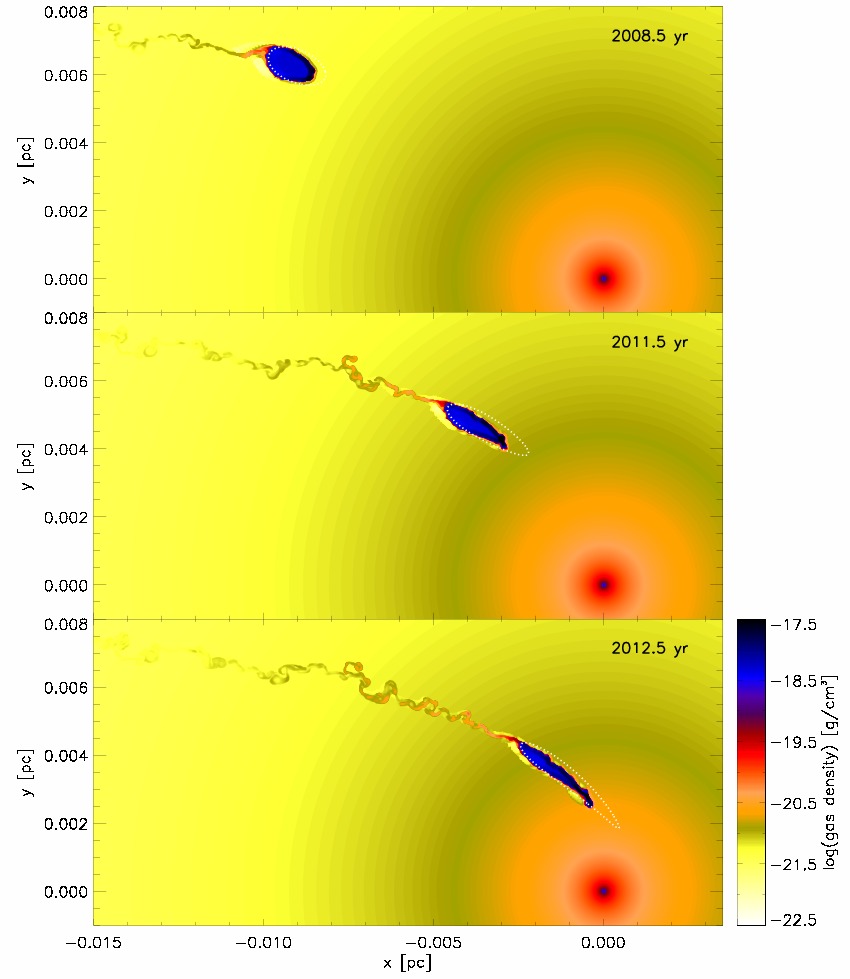}
\end{center}
\caption{Early phases of the evolution of a cold gas cloud (model CC01) with a mass equal to G2
that starts in pressure equilibrium with the surrounding as a spherical clump
in the year 1995. Colors correspond to the logarithm of gas density. White dotted contours
correspond to a test particle simulation where gas particles move on ballistic orbits.
\label{fig:four} }
\end{figure}

\begin{figure}[ht]
\begin{center}
\includegraphics[width=1.0\textwidth]{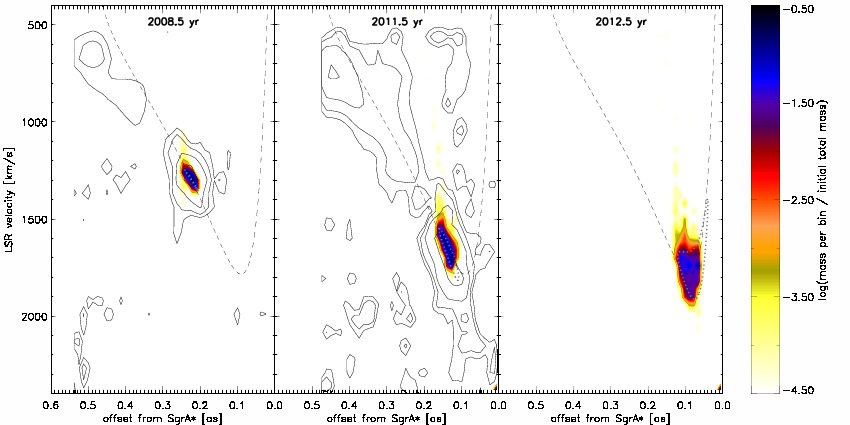}
\end{center}
\caption{Position versus line-of-sight velocity map of the "in-situ" formation model CC01
(see also Figure 2 of Gillessen et al. 2012), which corresponds to a cloud that forms in 1995.
The contours show the Br$\gamma$ emission of G2, obtained with SINFONI on the VLT (Gillessen et al. 2012).
Colors correspond to the total mass of gas as predicted by the simulation
in each position-velocity bin with the values given by the color bar. The dashed line shows the ballistic
orbit of a point particle in the gravitational potential of the SMBH, centered on G2. The dotted lines
show the distribution if each gas element of the cloud would move on a balistic orbit.
In 2008.5 the cloud is still compact. Over the next 3 years its develops a strong velocity shear.
Overall the calculations are in excellent agreement with the observations.
The predicted structure of G2 in the PV-diagram for the year 2012.5 is shown in the right panel.
\label{fig:five} }
\end{figure}

\begin{figure}[ht]
\begin{center}
\includegraphics[width=1.0\textwidth]{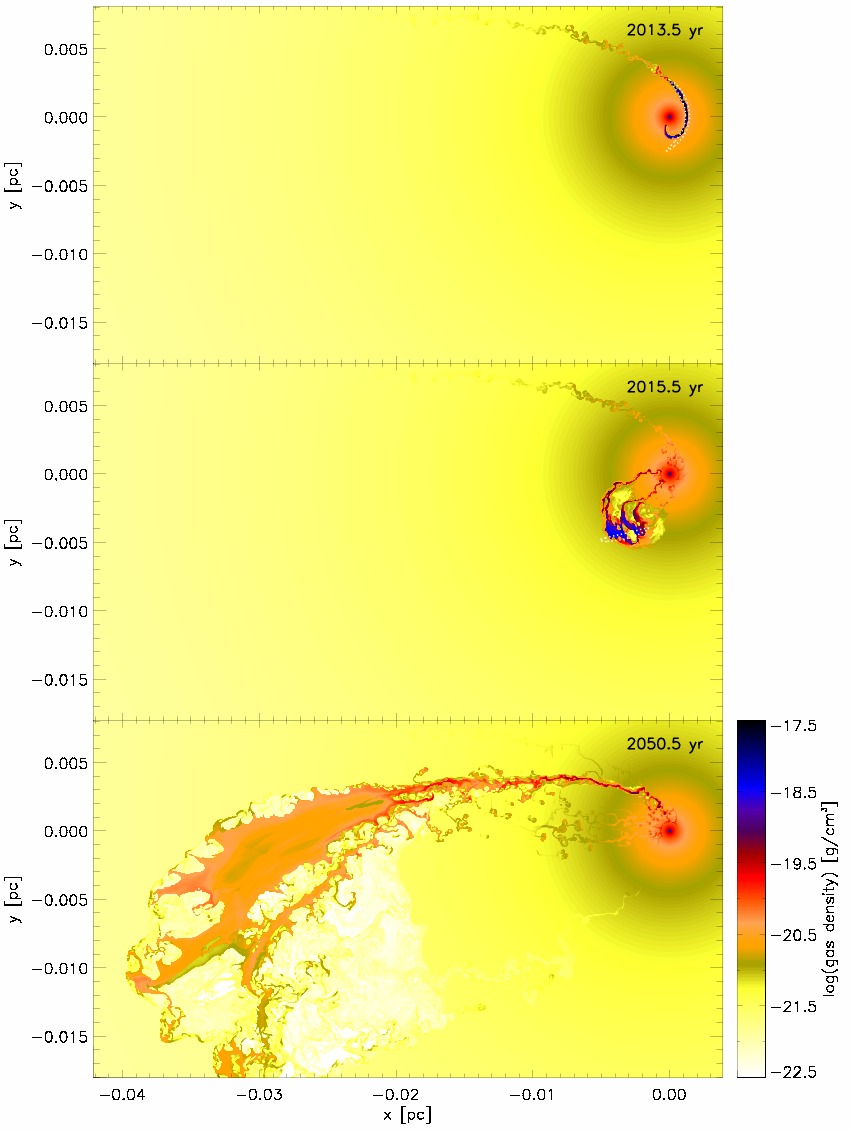}
\end{center}
\caption{Late phases of the evolution of model CC01.
In the year 2050 the cloud has broken up into a long filament that feeds Sgr A$^*$.
\label{fig:six} }
\end{figure}

\begin{figure}[ht]
\begin{center}
\includegraphics[width=1.0\textwidth]{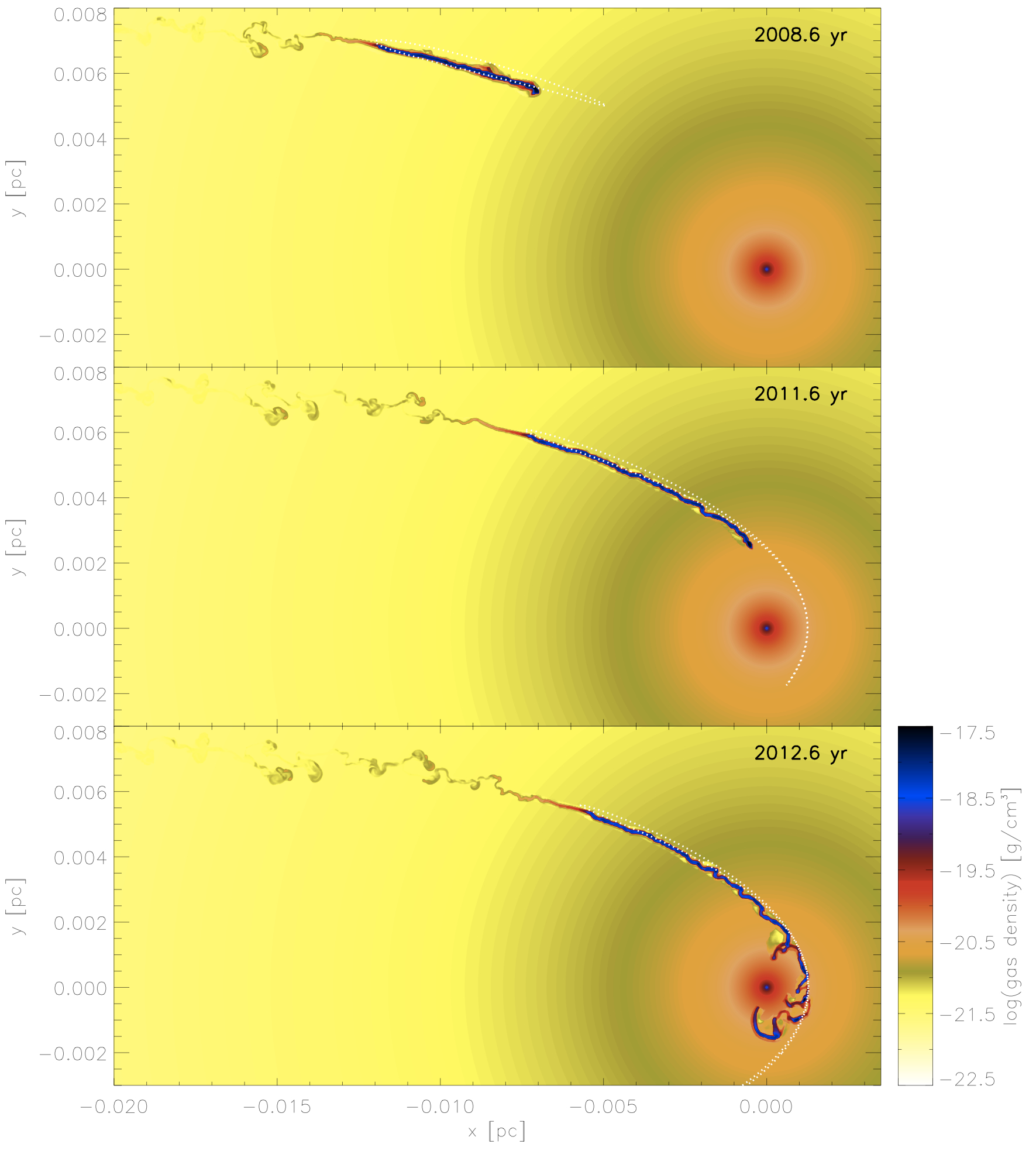}
\end{center}
\caption{Same as Figure 4 for model CC02. Now the cloud 
starts in pressure equilibrium at apocenter.
\label{fig:seven} }
\end{figure}

\begin{figure}[ht]
\begin{center}
\includegraphics[width=1.0\textwidth]{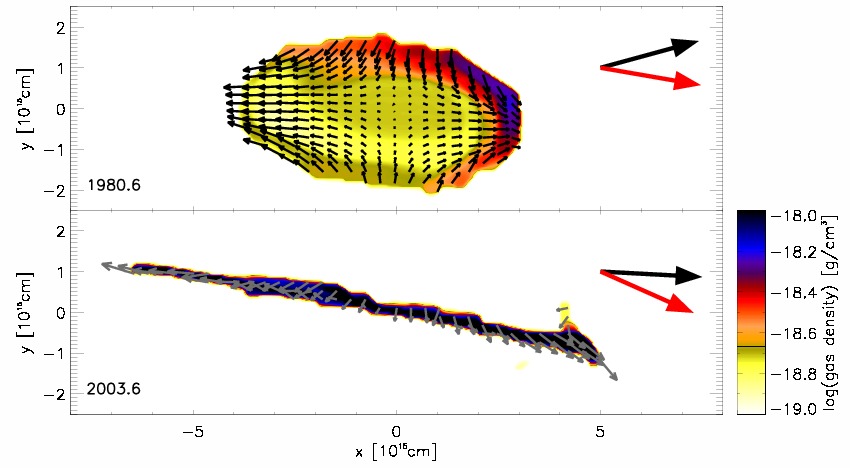}
\end{center}
\caption{Density distribution and velocity field in the cloud in model CC02 that starts
in pressure equilibrium at apocenter. The upper panel shows the early cloud structure in the
year 1980.6. The two large vectors show the radial direction towards the SMBH (red) and the
mean orbital velocity vector (black) of the cloud, respectively. The velocity field represents the 
internal velocity of the cloud's gas after subtracting the center-of-mass velocity (large
black arrow). The size of the velocity vectors scale linearly with velocity. As reference,
the large arrows in the upper and lower panel correspond to a velocity of 485 km/s and
1135 km/s, respectively.
The cloud is compressed perpendicular to its orbit and tidally sheared along its orbit.
Due to ram pressure a high-density shell is visible at the upstream front of the cloud.
The lower panel shows the structure in the year 2003.6. The cloud has been torn apart into
a long spaghetti-like filament.
\label{fig:eight} }
\end{figure}

\begin{figure}[ht]
\begin{center}
\includegraphics[width=1.0\textwidth]{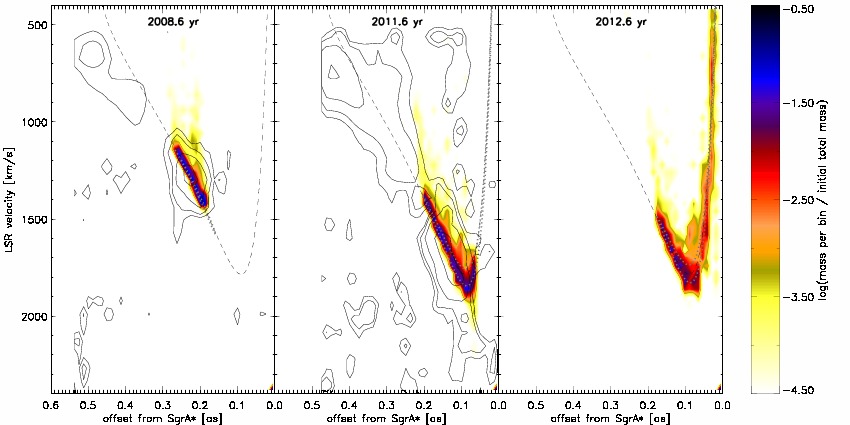}
\end{center}
\caption{Position-velocity diagram of model CC02, a cloud that starts at apocenter.
\label{fig:nine} }
\end{figure}

\begin{figure}[ht]
\begin{center}
\includegraphics[width=0.9\textwidth]{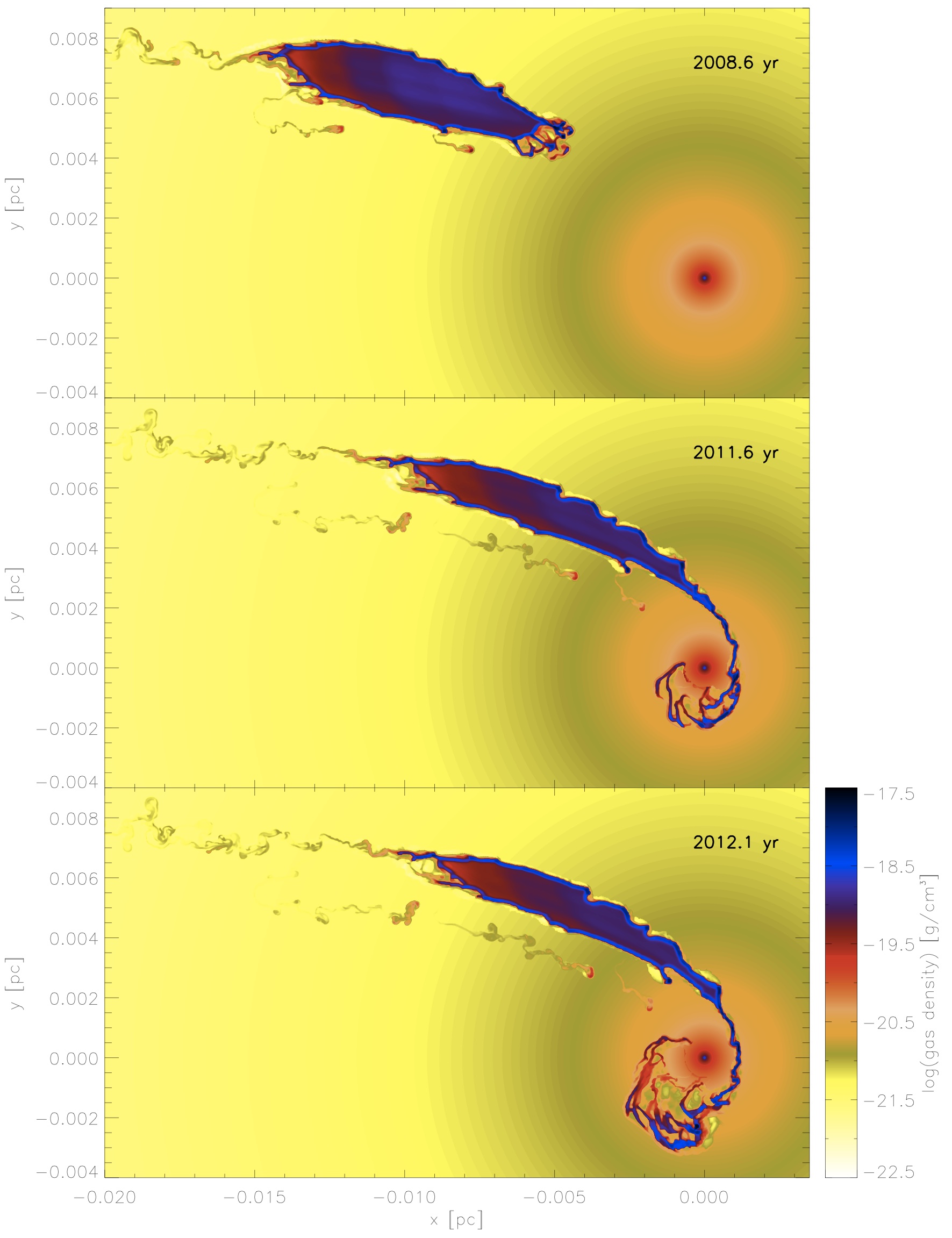}
\end{center}
\caption{Evolution of model CC03, a cloud that starts at apocenter with a density and pressure that
is a factor of 100 larger than its surrounding. Due to the resulting expansion the cloud evolves
into a large clump with an internal pressure that is significantly smaller than the surrounding.
\label{fig:ten} }
\end{figure}

\begin{figure}[ht]
\begin{center}
\includegraphics[width=1.0\textwidth]{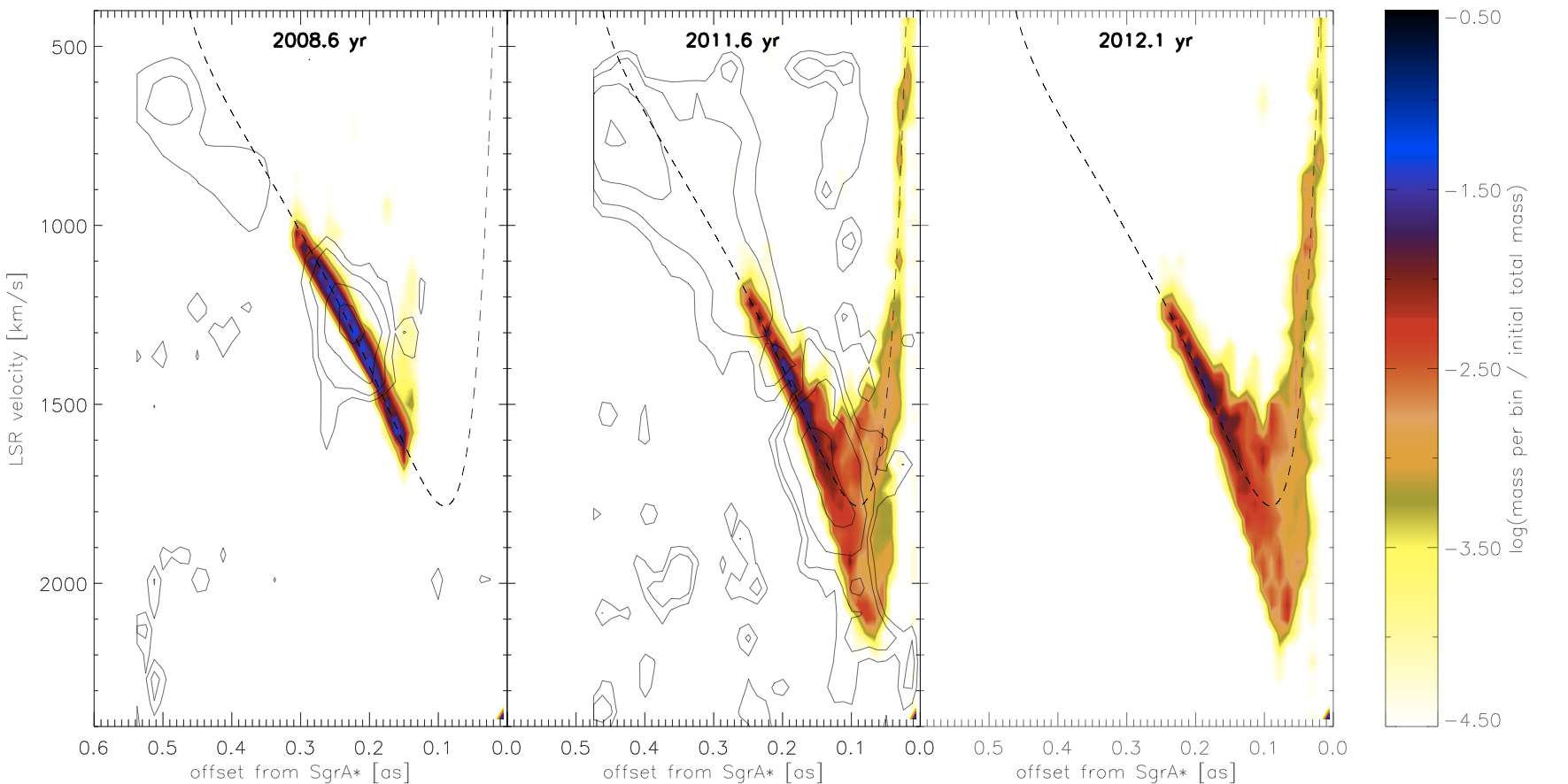}
\end{center}
\caption{PV-diagram of model CC03.
\label{fig:eleven} }
\end{figure}

\begin{figure}[ht]
\begin{center}
\includegraphics[width=1.0\textwidth]{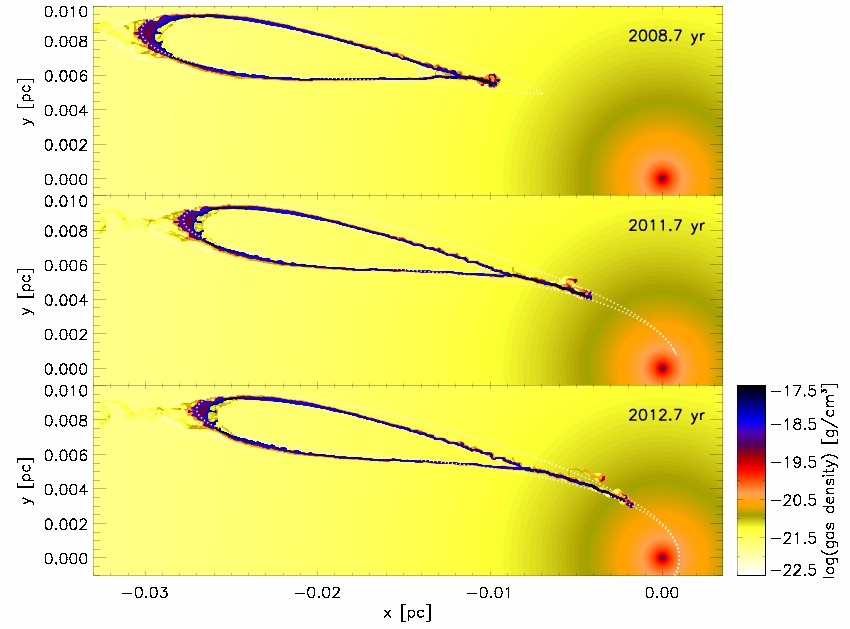}
\end{center}
\caption{Evolution of the spherical shell model SS01, which assumes that gas is initially distributed in a ring-like
structure at apocenter.
\label{fig:twelve} }
\end{figure}

\begin{figure}[ht]
\begin{center}
\includegraphics[width=1.0\textwidth]{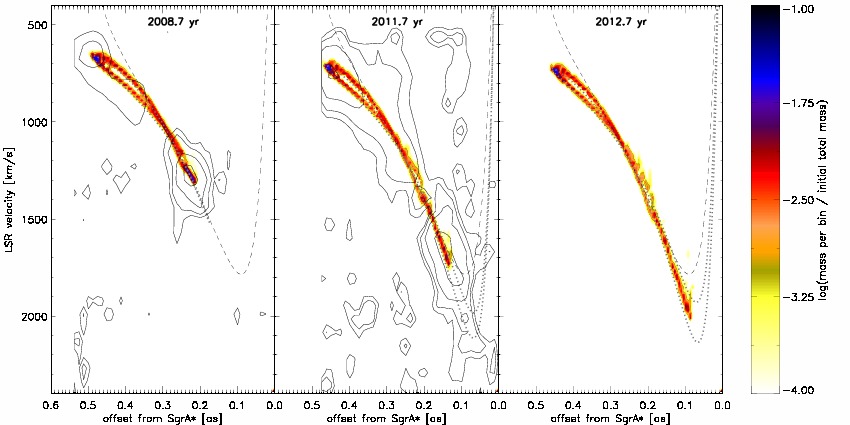}
\end{center}
\caption{PV-diagram of model SS01 which forms two bright regions of emission
with diffuse material in between, similar to the observations.
\label{fig:thirteen} }
\end{figure}

\noindent
{\bf Acknowledgments:}
The research of A.B. is supported by a Max Planck Fellowship and by the DFG
Cluster of Excellence ``Origin and Structure of the Universe''. We thank Chris McKee,
Eliot Quataert, Avi Loeb, Joachim Puls and Rolf Kudritzki for helpful discussions.

\end{document}